# New Calibrations and Time Stability of the Response of the INTERCAST CR-39

S. Cecchini[a,b], G. Giacomelli[a], M. Giorgini[a] *, L. Patrizii[a], P. Serra[a]

[a]Dip. di Fisica dell'Universita' di Bologna e Sezione INFN di Bologna,
Viale C. Berti Pichat 6/2, I-40127 Bologna, Italy

[b]CNR, Istituto TESRE, Via P. Gobetti 101, I-40129 Bologna, Italy

**Abstract**

We present new calibrations of different production batches (from 1989 to 1999) of the INTERCAST CR-39, using the BNL-AGS 1 A GeV iron beam. The comparison with previous results, obtained with the 158 A GeV lead beam from the CERN-SPS shows that, while each production batch has a different calibration curve (mainly due to minor differences in the production conditions), the aging effect is negligible. We also tested the dependence of the CR-39 response from the time elapsed between exposure and analysis (fading effect). The fading effect, if present, is less then 10%. It may be compatible with the experimental uncertainities on the bulk etching rate $v_B$.

**\* Corresponding author:**

Dr. Miriam Giorgini

Dipartimento di Fisica dell'Universita' e Sezione INFN di Bologna,

Viale C. Berti Pichat 6/2, I-40127 Bologna, Italy.

Tel : +39-51-2095235   ;   Fax : +39-51-2095269   ;   E-mail : miriam.giorgini@bo.infn.it

# 1. Introduction

The INTERCAST CR-39 is used as a nuclear track detector in the MACRO (MACRO Coll., 1993) and SLIM (SLIM Coll., 2000) experiments. MACRO is a large area detector, located in the Hall B of the underground Gran Sasso Laboratory. One of its primary purpose is the search for magnetic monopoles with various subdetectors, one of which exploits the 1200 m$^2$ nuclear track detector CR-39. SLIM is a 400 m$^2$ passive detector devoted primarely to a "light" magnetic monopole search; it is located at the Chacaltaya high altitude Laboratory at 5230 m above sea level; it is assumed that its exposure will last 4 years.

In both cases the CR-39 was produced in different batches with slightly different sensitivities. The long exposure times require the knowledge of aging and fading effects. The aging effect is the loss of capability of detection in the period between the production and the exposure. The fading effect is the partial or total loss of the damage due to a highly ionizing particle in the time elapsed from exposure to analysis. For the aging effect, we compared the response of CR-39 exposed to ion beams at different times after production and analyzed immediately after exposure. For the fading effect we compared the CR-39 response in sheets exposed at the same time to the same beam and analyzed after different periods.

To establish the response of the nuclear track detector CR-39 (calibration curves), stacks of CR-39 sheets were exposed to a 158 A GeV (A GeV means GeV/a.m.u.) Pb$^{82+}$ ion beam at the CERN-SPS in Dec. 94 (Lead94) and Dec. 96 (Lead96) and to a 1 A GeV Fe$^{26+}$ ion beam at BNL-AGS in Dec. 99 (Iron99). The typical thickness of a CR-39 sheet was ~ 1.4 mm. The range of the Fe$^{26+}$ and Pb$^{82+}$ ions in CR-39 is ~ 14 cm and some meters, respectively.

The CR-39 was manufactured by the INTERCAST Europe Co. of Parma (Italy) from 1989 to 1999 in different production batches, using the same chemical composition and the same curing cycle (MACRO Coll., 1991, 1993).

In each stack, the ion beam crossed a target (where the incident nuclei fragmented) and CR-39 sheets before and after the target. The thickness of the targets (~ 1 cm) was chosen in order to have about 50% of survived beam. Both the beam and the fragment ions with electric charges above the detector threshold were recorded. The exposures were carried out at normal incidence; the beam density was ~ 400 ions/cm$^2$. After exposure, the CR-39 foils were etched in a 6N NaOH water solution at 70 °C for 30 hours in the Pb cases and for 45 hours in the Fe case. During the etching, the CR-39 is removed at a bulk velocity $v_B$ and at a track velocity $v_T$, greater than $v_B$, along an ion path. Therefore, for ions with $Z/\beta$ greater than the detector threshold (5$e$ for the INTERCAST CR-39), the etching process leads to the formation of a conical etch-pit on each side of a detector foil along the ion path (Fleischer et al., 1975; Cecchini et al., 1996; Giacomelli et al., 1997, 1998). In the used etching conditions, the bulk etching rate $v_B$ for the INTERCAST CR-39 ranges between 1.092 and 1.231 µm/h.

The reduced etch rate $p = v_T/v_B$ is used to measure the detector response. It was determined using the relation (Fleischer et al., 1975)

$$p = [1+(D/2v_B t)^2] / [1-(D/2v_B t)^2] \qquad (1)$$

For normally incident particles, $D$ is the diameter of the etch-pit cone base. Previous calibrations had shown that the reduced etch rate $p$ is a function of the Restricted Energy Loss (REL) only (Fleischer et al., 1975; Cecchini et al., 1996). For each detected charge, the REL was computed using the Bethe-Block formula (Particle Data Group, 2000) with energy cut-off at 200 eV and is a function of $(Z/\beta)^2$. This means that once an ion has been identified by its base cone area (or diameter), we know its $p$-value and REL. We checked that the REL values on a CR-39 sheet were constant for Pb and Fe ions; in these conditions the tracks have conical shapes.

The track diameters were obtained from the base cone area measurements made with the Elbek automatic image analyzer system (Noll et al., 1988). About 50000 tracks were measured on each foil. Fiducial marks and patterns of tracks were used to track the path of each detected ion along the stack. The $v_B$ was determined via thickness measurements before and after the etching procedure.

Each stack was composed of CR39 foils from different batches; it was then possible to determine the response of each batch to the same incident ion.

For each batch we measured three consecutive foils (6 faces) for the Lead96 calibrations, two consecutive foils (4 faces) for the Lead94 calibrations, one foil (2 faces) for the Lead94 calibrations used for studying fading effects and for the Iron99 calibrations. A "signal" was defined as the spatial coincidence of similar base cone areas in the required number of faces; the average of the measured base areas was computed.

A typical average base area distribution is shown in Fig. 1a for the Lead96 beam. It refers to CR-39 made in Nov. 95 and etched for 30 h in 6N NaOH at 70 °C shortly after exposure. A large number of peaks is evident. Each peak corresponds to a nuclear fragment with different charge. Due to the worsening of the charge resolution with increasing charge and the relatively small number of measured faces, only fragments with charge up to $Z \sim 35e$ are clearly separated.

At the first peak the charge $Z = 5e$ was assigned by comparison with our previous calibrations (Cecchini et al., 1993, 1996) and assuming that the threshold of the CR-39 was the same as for samples of the same age at the time of exposure and etched under the same conditions.

Fig. 1b refers to CR-39 exposed to the iron beam, etched for 45 h in 6N NaOH at 70 °C (2 measured faces). The CR-39 used was made in Dec. 96. Here all the nuclear fragments with charges from $6e$ to $26e$ are present. Due to some background not completely removed using only two faces, we had to cut low values of the averaged area loosing the 5*e* ion signal.

## 2. Calibrations with the Iron Beam

In Fig. 2 are shown the calibration curves (*p* vs REL) for various CR-39 production batches since Apr. 89 to Apr. 99 using the BNL-AGS $Fe^{26+}$ beam of 1 A GeV. The Apr. 99 CR-39 was exposed with an aging time of about 8 months, so its calibration curve is particularly high (Cecchini et al., 1996). As far as the other curves we can say that each production batch has its own calibration curve with no evident connection with the production time (see for istance the calibration curves of Nov. 89 and Nov. 97 production batches that are very similar in spite of their really different production ages). The variation is mainly due to not perfectly identical conditions during the production. Further studies of these effects are in progress.

## 3. Comparison of the Calibrations with Iron and Lead Beams

We have compared the calibration curves of CR-39 from the same production batches, exposed to the CERN-SPS lead beam (Lead96) and the BNL-AGS iron beam (Iron99) to test the reproducibility of the calibrations and the aging effect. Fig. 3 shows the results for the Nov. 89 and Nov. 95 production batches. Within errors no effect is visible.

## 4. Fading Effects in CR-39

In order to test the dependence of the CR-39 response from the time elapsed between exposure and etching (fading effect), several sets of samples belonging to the same production batch were exposed in Dec. 94 to the 158 A GeV lead ion beam and etched after different times. In Fig. 4 the results for two sets etched 0.5 y and 4.5 y after the exposure, respectively, are shown. The response of the samples etched 4.5 y after exposure is less than 10% lower than that for the samples etched shortly after exposure. The difference is within our experimental uncertainties on bulk velocity, shown as bands in Fig. 4. That means that our detectors after 4.5 y are still well active.

## 5. Conclusions

The experimental results quoted above indicate that the response of the INTERCAST CR-39 used in the MACRO and SLIM experiments shows a fair time stability. There are no indications of aging effects for material older than 10-12 months. From the study of fading effects we can be confident that, up to now, the detectors are at full sensitivity.

## Acknowledgments

We gratefully acknowledge the staff of the CERN-SPS and BNL-AGS for the help during the exposures. We acknowledge the collaboration of E. Bottazzi, L. Degli Esposti, D. Di Ferdinando, V. Togo, C. Valieri and the general services of the Bologna Section of INFN.

## References

,
Cecchini S., Dekhissi, H., Giacomelli, G., Katsavounidis, E., Margiotta, A.R., Patrizii, L., Pedrieri, F., Serra, P. and Spurio, M., 1993. Calibration of the INTERCAST CR-39, Nucl. Tracks Radiat. Meas. 22, 555-558.

Cecchini, S., Dekhissi, H., Garutti, V., Giacomelli, G., Katsavounidis, E., Mandrioli, G., Margiotta, A.R., Patrizii, L., Popa, V., Serra, P., 1996. Calibration with relativistic and low velocity ions of a CR-39 nuclear track detector, Il Nuovo Cimento 109 A, 1119-1128.

Fleischer, R., Price, P.B. and Walker, R.M., 1975. Nuclear Tracks in Solids, University of California Press.

Giacomelli, G., Patrizii, L., Popa, V., Serra, P. and Togo, V., 1997. New results from exposures of CR-39 Nuclear Track Detectors, Nucl. Tracks Radiat. Meas. 28, 217-222.

Giacomelli, G., Giorgini, M., Mandrioli, G., Manzoor, S., Patrizii, L., Popa, V., Serra, P., Togo, V. and Vilela, E.C., 1998. Extended calibration of a CR-39 nuclear track detector with 158 A GeV Pb ions, Nucl. Instrum. Meth. A 411, 41-45.

MACRO Coll., 1991. Improvements of the CR-39 polymer for the MACRO experiment at the Gran Sasso Laboratory, Nucl. Tracks Radiat. Meas. 19, 641-646.

MACRO Coll., 1993. First supermodule of the MACRO detector at Gran Sasso, Nucl. Instrum. Meth. A 324, 337-362.

Noll, A., Rusch, G., Rocher, H., Dreute, J. and Heinrich, W., 1988. The Siegen automatic measuring system for nuclear track detectors: new developments, Nucl. Tracks Radiat. Meas. 15, 265.

Particle Data Group, 2000. Review of Particle Physics, Eur. Phys. J. C 15, 165.

SLIM Coll., 2000. Search for "light" Magnetic Monopoles, (SLIM proposal) hep-ex/0003028.

**Figure Captions**

Fig. 1. (a) Base cone area distribution for the CR-39 exposed to the CERN $Pb^{82+}$ beam of 158 A GeV; the CR-39 was etched in 6N NaOH at 70 °C for 30 hours. The base areas were averaged over 6 faces. (b) Base cone area distribution for the CR-39 exposed to the $Fe^{26+}$ beam of 1 A GeV; the CR-39 was etched in 6N NaOH at 70 °C for 45 hours. The base areas were averaged over 2 faces.

Fig. 2. Calibration ($p$ vs Restricted Energy Loss) curves for various CR-39 production batches exposed to the 1 A GeV iron beam (BNL-AGS, 1999).

Fig. 3. Calibrations of CR-39 from the same production batch exposed to the CERN-SPS lead beam and to the BNL-AGS iron beam with different aging times.

Fig. 4. Calibrations of CR-39 from the same production batch exposed to the CERN-SPS lead beam and analysed 0.5 y and 4.5 y after exposure.

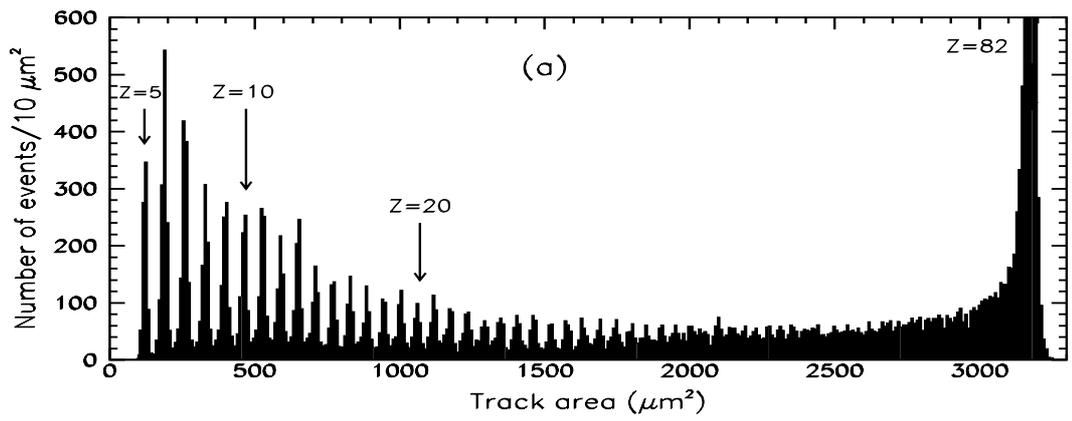

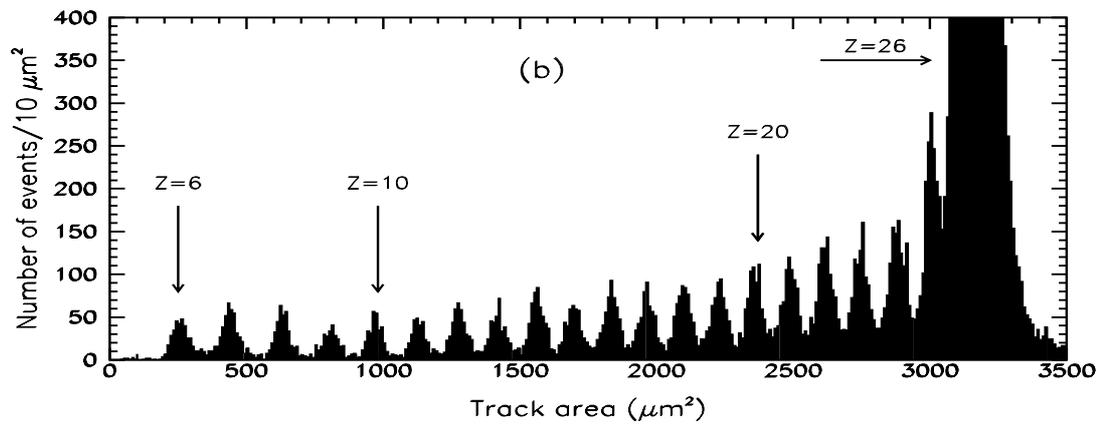

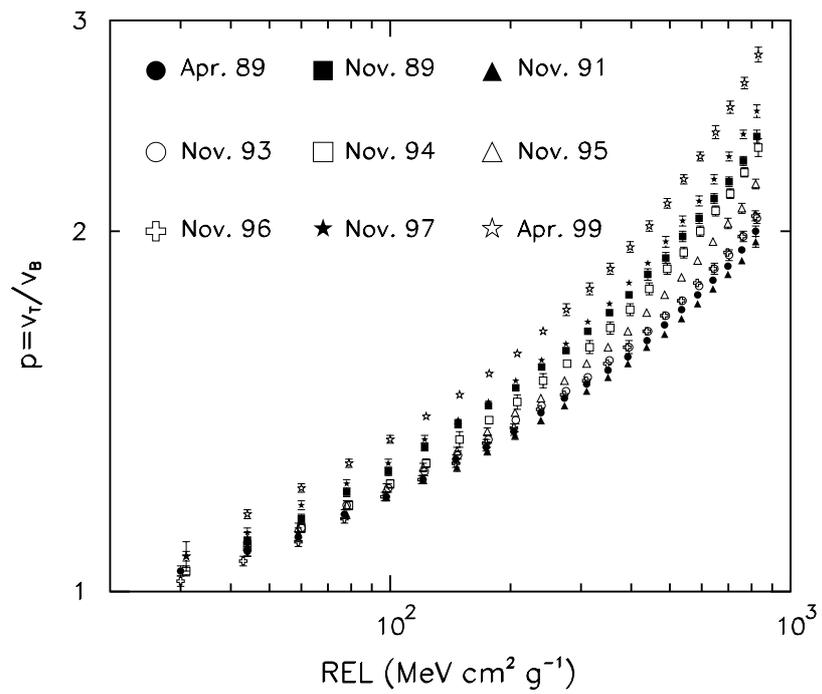

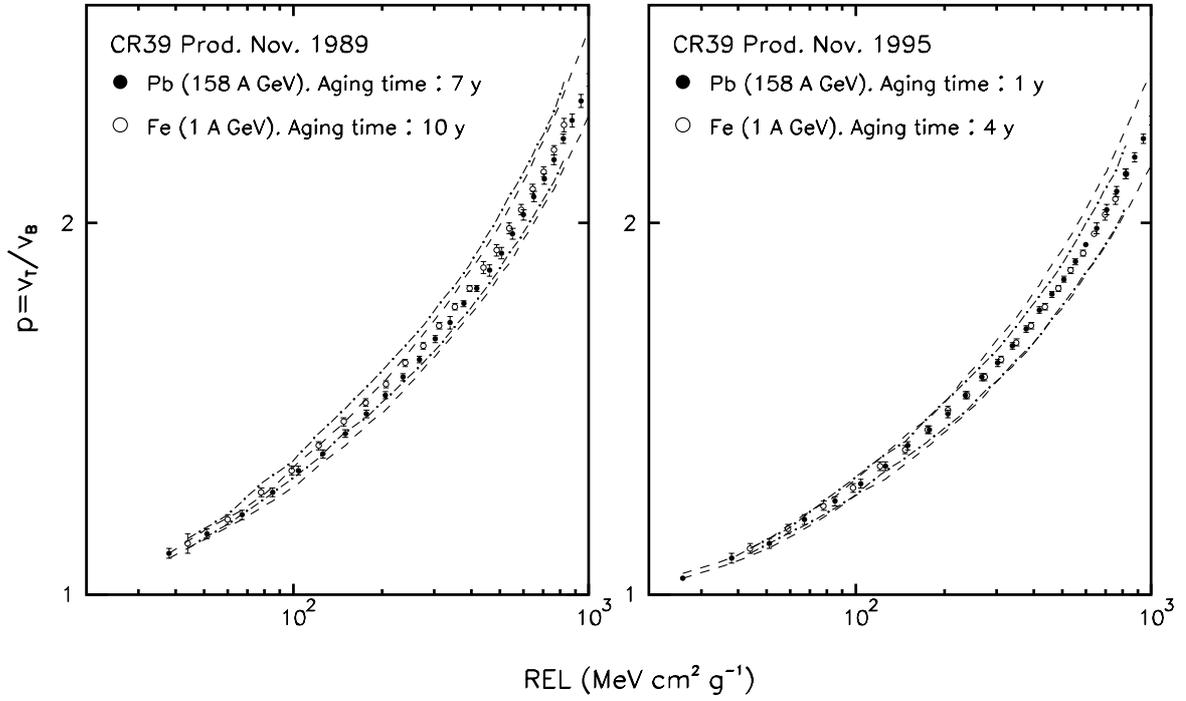
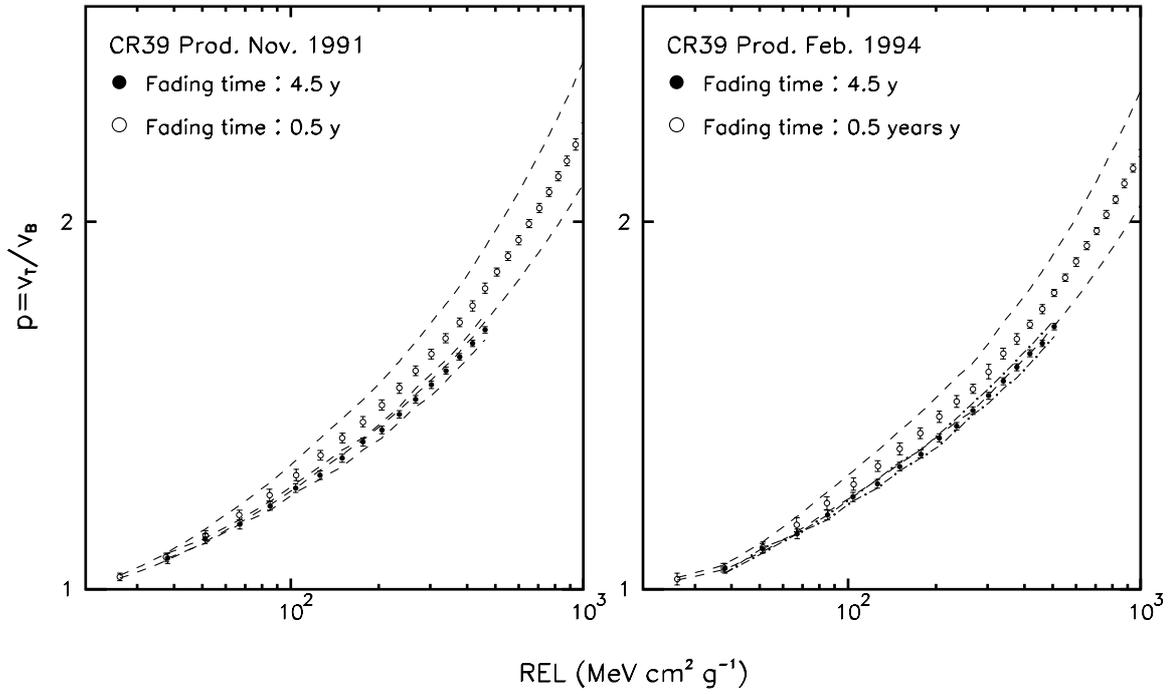